\documentclass[prl,showpacs,twocolumn,a4paper]{revtex4}
\usepackage{graphics}

\newcommand{\centreline}{\centerline}

\newcommand{\Rcm}{cm\ensuremath{^{-1}}} 
\newcommand{\Tc}{\ensuremath{T_{c}}}
\newcommand{\caxis}{\textit{c}-axis}
\newcommand{\abplane}{\textit{ab}-plane}
\newcommand{\acface}{\textit{ac}-face}
\newcommand{\zz}{\ensuremath{zz}}

\newcommand{\xz}{\ensuremath{xz}}
\newcommand{\xx}{\ensuremath{xx}}
\newcommand{\xy}{\ensuremath{xy}}

\begin{document}

\preprint{Version 3}

\title{\caxis\ Raman Scattering in {MgB$_{2}$}:\\
       Observation of a Dirty-Limit Gap in the $\pi$-bands}

\author{J. W. Quilty}
\author{S. Lee}
\author{S. Tajima}
\affiliation{Superconductivity Research Laboratory, 
             International Superconductivity Technology Center,\\
             1-10-13 Shinonome, Koto-ku, Tokyo 135-0062, Japan}
\author{A. Yamanaka}
\affiliation{Chitose Institute of Science and Technology, Chitose, 
             Hokkaido 066-8655, Japan}

\begin{abstract}

Raman scattering spectra from the \acface\ of thick MgB$_{2}$ single
crystals were measured in \zz, \xz\ and \xx\ polarisations. In \zz\
and \xz\ polarisations a threshold at around 29~\Rcm forms in the
below \Tc\ continuum but no pair-breaking peak is seen, in contrast to
the sharp pair-breaking peak at around 100~\Rcm\ seen in \xx\
polarisation. The \zz\ and \xz\ spectra are consistent with Raman
scattering from a dirty superconductor while the sharp peak in the
\xx\ spectra argues for a clean system. Analysis of the spectra
resolves this contradiction, placing the larger and smaller gap
magnitudes in the $\sigma$ and $\pi$ bands, and indicating that
relatively strong impurity scattering is restricted to the $\pi$
bands.

\end{abstract} 


\maketitle

MgB$_{2}$ exhibits a rich multiple-band structure which has been
observed by a number of experimental techniques including
angle-resolved photoemission spectroscopy, de Haas-van Alphen effect
and Hall resistivity
\cite{Uchiyama02:prl88,Yelland02:prl88,Eltsev02:prb66}. These
results confirm band structure calculations
\cite{Kortus01:prl86,An01:prl86} and reveal strongly two-dimensional
(2D) $sp_{x}p_{y}$ ($\sigma$) bands and three-dimensional (3D) $p_{z}$
($\pi$) bands. The existence of different superconducting gaps
associated with the different band systems has, in the time since its
theoretical proposal \cite{Shulga01:0103154,Liu01:prl87}, become the
foremost topic of discussion surrounding MgB$_{2}$. While observations
of two gaps in polycrystal samples have been reported (see
Ref.~\onlinecite{Buzea01:sst14} for a review), it is extremely
difficult to distinguish between a strongly anisotropic single gap and
two distinct gaps from polycrystal measurements. Moreover, none of the
reports could conclusively associate the gaps with particular band
systems because they could not distinguish the $\sigma$ and $\pi$
bands.

Raman scattering, which has the benefits of excellent energy
resolution, a relatively large penetration depth and the ability to
selectively measure different portions of the Fermi surface(s), has
been somewhat inconsistent on the subject of multiple gaps in
MgB$_{2}$. Recent single crystal
measurements restricted to \abplane\ polarised spectra showed only a
single superconducting gap \cite{Quilty02:prl88}, while data from
polycrystalline MgB$_{2}$ were interpreted in terms of two
superconducting gaps both residing on the $\sigma$-bands
\cite{Chen01:prl87}. It was suggested that the second gap feature
might appear only in out-of-plane polarisations \cite{Quilty02:prl88}
and a complete set of polarisation data would permit superconductivity
on the $\sigma$ and $\pi$ Fermi surfaces to be discerned
\cite{Quilty03:pc385}. Thus, to resolve the inconsistency between Raman
measurements and to clarify the multiple gap issue in MgB$_{2}$,
out-of-plane measurements of single crystals are crucial.

\begin{figure}
    \centreline{\includegraphics{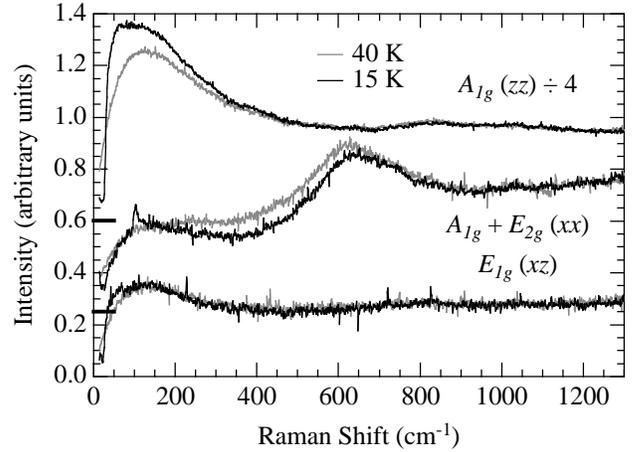}}
    \caption{Raman spectra at 40 and 15~K from the \acface\ of a
        single crystal of MgB$_{2}$. The broad $E_{2g}$ phonon at
        620~\Rcm\ appears only in \xx\ polarisation. Similarly, the
        electronic continuum is strongly polarisation dependent.}
    \label{fig:PolarisationDependence}
\end{figure}
In this letter we report \acface\ Raman scattering measurements from
thick MgB$_{2}$ single crystals with {\acface} dimensions 100--150 by
100--150~$\mu\mathrm{m}$ and superconducting transition temperatures
of 38.0--38.4~K. Raman measurements were performed in
near-backscattering configuration with polarisation geometries \zz
($A_{1g}$), \xx\ ($A_{1g} + E_{2g}$) and \xz\ ($E_{1g}$); details of
crystal fabrication and experimental setup are reported elsewhere
\cite{Quilty02:prl88,Lee01:jpsj70,Quilty03:pc385}. The Raman vertex
$\gamma(\mathbf{k})$, which weights the electronic Raman scattering
cross section
\cite{Klein84:prb29,Devereaux92:prb45,Devereaux96:prb54}, is a
function of the band energies; the anisotropy of the $\sigma$ and
$\pi$ bands gives $\gamma_{\pi}(\mathbf{k}) \neq 0$ in \zz\ and \xx\
polarisations while $\gamma_{\sigma}(\mathbf{k}) \approx 0$ for \zz\
\cite{Quilty03:pc385}. Thus \zz\ polarised spectra
isolate the $\pi$ band response while both $\sigma$ and $\pi$ bands
contribute to \xx\ polarisation.
	 
Raman spectra taken at 40 and 15~K from one of the single crystals
studied are shown in Fig.~\ref{fig:PolarisationDependence} for three
different polarisation conditions in the \acface. All spectra have
been corrected for the Bose thermal contribution. The polarisation
dependence of the the $E_{2g}$ symmetry phonon, seen at around
620~\Rcm\ in the \xx\ spectrum, confirms both the crystal alignment in
our experiment and, incidentally, micro-Raman results on single
crystallites \cite{Hlinka01:prb64}.

A strong electronic continuum is present in all spectra, constituting
the principal component of the \zz\ and \xz\ spectra and always
showing greater intensity in \zz\ polarisation. Notably, the
superconductivity-induced renormalization of the continuum seen at low
frequencies in all polarisations has markedly different character in
\zz\ and \xz\ compared to \xx\ polarisation. As expected, the \xx\
continuum at 15~K exhibits the sharp pair-breaking peak at around
100~\Rcm\ seen in the same polarisation measured in-plane
\cite{Quilty02:prl88}. Meanwhile the \zz\ and \xz\ renormalizations
are very similar in character, showing a shoulder around 25--40~\Rcm\
accompanied by an increase in intensity above the threshold relative
to the normal-state continuum. Despite the formation of a threshold,
no sharp coherence peak due to the breaking of Cooper pairs is seen in
the \zz\ (or \xz) polarised continua --- neither directly above the
threshold edge nor in the vicinity of 100~\Rcm. Similar temperature
dependence of the low-frequency continuum was also seen in spectra
from another six crystals (not shown here). Indeed, we have been
unable to find a sharp coherence peak anywhere in the \zz\ (or \xz)
spectra of any single crystals studied. Since the \zz\ and \xz\
polarisations are similar, the following results and discussion will
focus on the \zz\ continuum.

\begin{figure}
    \centreline{\includegraphics{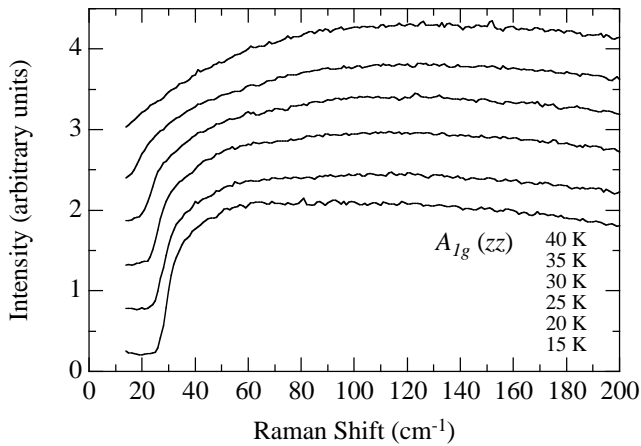}}
    \caption{Raman spectra from 40--15~K in \zz\ polarisation showing 
             the development of the threshold in the superconducting
             state. Successive spectra have been offset by 0.5 units
             along the vertical axis.}
    \label{fig:TempDep_Super_zz}
\end{figure}
Figure~\ref{fig:TempDep_Super_zz} shows the temperature dependence of
the \zz\ polarised electronic Raman continuum in the range 40--15~K.
As the temperature is reduced below \Tc\ a threshold forms in the
continuum, gaining a maximum extent at 15~K of around 29~\Rcm. As best
as can be discerned, the onset temperature appears to coincide with
\Tc\ and the temperature dependence appears to be BCS-like
\cite{Quilty03:pc385}. Recalling that \zz\ polarisation selects Raman
scattering from the $\pi$ Fermi surface only, this
superconductivity-induced renormalisation must be directly
related to the superconducting gap on the $\pi$ sheets. The residual
intensity below the shoulder edge varies between crystals and it is
likely that specular scattering from surface defects is the
predominant contribution.

\begin{figure}
    \centreline{\includegraphics{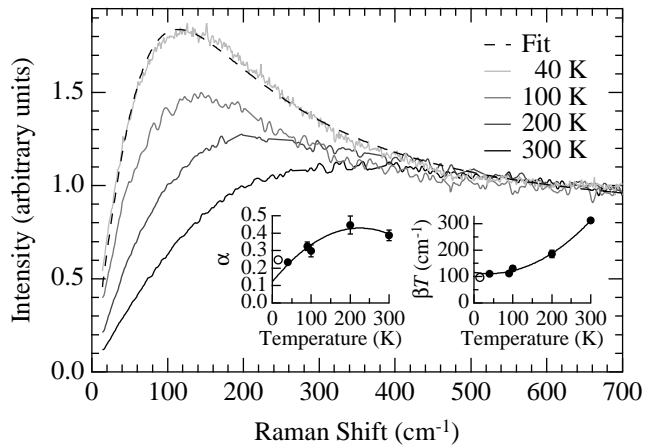}}
    \caption{Raman spectra from 300--40~K in \zz\ polarisation. The 
    dashed line represents a fit to the continuum, described in the
    text. Inset: the temperature dependence of the continuum fit
    parameters. The solid lines are guides for the eye.}
    \label{fig:TempDep_Normal_zz}
\end{figure}
In order to properly evaluate the superconductivity-induced spectral
change seen in Fig.~\ref{fig:TempDep_Super_zz} we must first examine
the normal state electronic Raman continuum. As shown in
Fig.~\ref{fig:TempDep_Normal_zz}, the \zz\ continuum is well described
over the entire frequency and temperature range by a relaxational
response due to impurity scattering \cite{Devereaux92:prb45}
\begin{equation}
    I(\omega) = \frac{A \omega \Gamma}
                {\omega^{2} + \Gamma^{2}}\ ,
    \label{eq:nFL_continuum}
\end{equation}
where $A$ is an arbitrary intensity factor, $\Gamma$ is the impurity
scattering rate for each scattering channel, and the Raman vertex has
been absorbed into $A$. Anisotropy of the impurity scattering rate
will be revealed in the polarisation-dependence of the Raman continua
\cite{Devereaux92:prb45}. To accommodate the nearly
flat continuum at higher frequencies a frequency-dependent impurity
scattering rate $\Gamma(\omega,T)~=~\sqrt{(\alpha\omega)^2 + (\beta
T)^2}$ is introduced, were $\alpha$ and $\beta$ are constants of the
order of unity. An example fit to the 40~K spectrum is shown in
Fig.~\ref{fig:TempDep_Normal_zz} and the $\alpha$ and $\beta$
parameters are plotted against temperature (inset). Below 90~K the
static-limit impurity scattering rate $\Gamma_{\pi}(\omega \rightarrow
0)$ is around 110~\Rcm. These results indicate that the increase in
intensity above the threshold in the superconducting state \zz\ and
\xz\ continua is at least partly attributable to the trend of
increasing spectral weight seen in the normal state, and does not
represent a broad pair-breaking peak. The peakless threshold in the
superconducting state continuum remains to be explained.

A threshold \textit{sans} peak may be seen in the Raman spectrum of
some clean superconductors when the optical penetration depth $\delta$
is less than the superconducting coherence length $\xi$
\cite{Klein84:prb29}. To examine this possibility in MgB$_{2}$, we use
$\delta = \sqrt{2/{\mu_{0}\sigma\omega}}$ and take $\sigma = 0.18
\times 10^{6}~\Omega^{-1}\mathrm{m}^{-1}$ at $\omega = 514.5$~nm
\cite{Kuzmenko02:ssc121} to obtain $\delta \approx 500$~\AA. In
comparison, estimates of $\xi_{c}$ range from 23~\AA\ -- 50~\AA\
\cite{Eltsev02:prb65,deLima01:prl86} up to $\xi_{c} \approx 500$~\AA\
\cite{Eskildsen03:pc385}, giving $\delta \gtrsim \xi$. A sharp
pair-breaking peak is thus expected to appear in the absence of
significant impurity scattering
\cite{Klein84:prb29,Devereaux92:prb45}. The analysis of the
normal-state spectra indicates, however, that the \zz\ continuum
arises from relatively strong impurity scattering.

Devereaux \cite{Devereaux92:prb45} has developed a theory of Raman
scattering in dirty superconductors for the case of non-magnetic
impurity scattering. The channel-dependent Raman intensity is
expressed as
\begin{widetext}
\begin{equation}
    I(\omega) = 
        \frac{A\omega\Gamma}{\omega^{2} + \Gamma^{2}}
	\Theta(\omega - 2\Delta)\frac{4\Delta}{\omega+2\Delta}
        \left(
            \frac{(\omega-2\Delta)^{2}}{4\Delta\omega}E(a) +
            \frac{\omega^{2} + \Gamma^{2} + 4\Delta}
                 {\omega^{2} + \Gamma^{2} + 2\Delta}K(a) +
            \frac{2(2\Delta\omega)^{2}}
                 {(\omega^{2} + \Gamma^{2})^{2} - (2\Delta\omega)^{2}}
                 \Pi(n,a)
        \right),
    \label{eq:DevereauxEq28}
\end{equation}
\end{widetext}
where $\Theta(x)$ is the theta function (0 for $x < 0$; 1
for $x > 0$), $\Delta$ is the magnitude of the superconducting gap and
$K$, $E$ and $\Pi$ are complete elliptic integrals of the first,
second and third kind. The arguments of the elliptic integrals are
\begin{eqnarray*}
    a & = & \frac{\omega - 2\Delta}{\omega + 2\Delta} \\
    n & = & \frac{(\omega - 2\Delta)^{2} + 
    (a\Gamma)^{2}\left( 1 - \left(\frac{2\Delta}{\omega\Gamma}\right)^{2} \right)}
    {(\omega - 2\Delta)^{2} + 
    \Gamma^{2}\left( 1 - \left(\frac{2\Delta}{\omega\Gamma}\right)^{2} \right)}.
\end{eqnarray*}
In the limit $\Gamma \rightarrow 0$ Eq.~(\ref{eq:DevereauxEq28})
yields a spectral line shape analogous to that for clean
superconductors \cite{Klein84:prb29}. Note that this model neglects
corrections to the Raman vertex (such as final state interactions) and
assumes an isotropic \textit{s}-wave gap \cite{Devereaux92:prb45}.
Despite indications that the gaps which form on the $\sigma$ and $\pi$
Fermi surfaces will be anisotropic in clean MgB$_{2}$
\cite{Choi02:nat418}, the use of an isotropic gap function is
justified by the expectation that gap anisotropy will be strongly
suppressed by impurity scattering \cite{Mazin02:0212417}.

\begin{figure}%[t]
    \centreline{\includegraphics{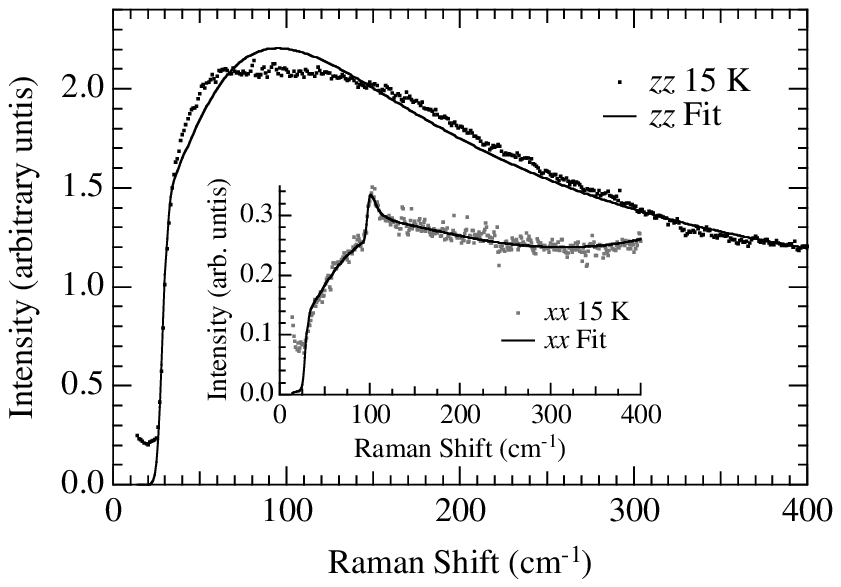}}
    \caption{The \zz\ polarised Raman spectrum at 15 K. The solid line
             shows the result of fitting to a single gap model in the
             presence of impurity scattering. Inset: The \xx\
             polarised Raman spectrum at 15 K, fitted with a two gap
             model.}
    \label{fig:SuperFit}
\end{figure}

Figure~\ref{fig:SuperFit} shows the \zz\ polarised Raman spectrum at
15~K. Using Eq.~(\ref{eq:DevereauxEq28}) with the frequency-dependent
impurity scattering rate defined above, and convolving with a Gaussian
to account for broadening effects
\cite{Klein84:prb29,Devereaux92:prb45}, the fit shown as the solid
line in Fig.~\ref{fig:SuperFit} is obtained. The agreement between the
fit and spectrum is adequate. The scattering rate obtained from the
fit $\Gamma_{\pi}(\omega \rightarrow 0, 15~K) = 100$~\Rcm\ is large
enough to suppress the pair-breaking peak, leaving only a threshold at
$2\Delta_{\pi} = 29$~\Rcm, consistent with the observed spectra. The
smearing Gaussian half-width was 3.5~\Rcm. We mention in passing that
the $\alpha$ and $\beta$ parameters from this fit at 15~K agree
reasonably well with extrapolated values from the normal state fits
(included in the inset to Fig.~\ref{fig:TempDep_Normal_zz} as open
circles).

The \xx\ polarised spectrum may also be fitted with
Eq.~(\ref{eq:DevereauxEq28}). Taking the simplest approach,
decomposing the electronic continuum into $\sigma$ and $\pi$
components and using Eq.~(\ref{eq:DevereauxEq28}) to describe each, a
satisfactory fit is obtained as shown in the inset to
Fig.~\ref{fig:SuperFit}. The phonon was represented by a Fano
lineshape and a linear term added to accommodate the weak luminescence
background. Additionally, $2\Delta_{\pi}$ was fixed at 29~\Rcm\ and
the smearing Gaussian width fixed at 3.5~\Rcm\ during the fits. Since
contributions from the different scattering channels have been
ignored, the fit results are channel-averaged. We obtain
$2\Delta_{\sigma} \sim 100$~\Rcm, $\Gamma_{\sigma} \sim 30$~\Rcm\ and
$\Gamma_{\pi} \sim 120$~\Rcm\ from the \acface\ \xx\ spectra of this
and another \cite{Quilty03:pc388} crystal. Fits to the \xx\ and \xy\
polarised spectra measured in-plane \cite{Quilty02:prl88} give
$\Gamma_{\sigma} \sim 20$~\Rcm\ and $\Gamma_{\pi} \sim 190$~\Rcm\ in
both polarisations. These results strongly suggest that anisotropy in
the $\sigma$ and $\pi$ impurity scattering rates, if present, is not
pronounced.

Our analysis thus indicates that below \Tc\ a dirty-limit gap forms on
the $\pi$ sheets and a clean-limit gap forms on the $\sigma$ sheets in
MgB$_{2}$. Different impurity scattering rates in the $\sigma$ and
$\pi$ bands have been reported in other experiments
\cite{Kuzmenko02:ssc121} and calculations \cite{Mazin02:prl89}, and
our measurements of the static-limit $\Gamma_{\sigma}$ and
$\Gamma_{\pi}$ agree within a factor of two with those deduced from
\abplane\ and \caxis\ resistivity measurements of similar crystals
\cite{Eltsev02:prb66}. We note that the theoretical expectation for
this two-band system is that $\sigma$-$\pi$ interband scattering will
be much smaller than intraband scattering
\cite{Golubov97:prb55,Mazin02:prl89}. Based on the presence of a
single pair-breaking peak and clean-limit superconductivity in
MgB$_{2}$, we previously noted only a single gap in \xx\ and \xy\
spectra \cite{Quilty02:prl88}. Here we find that two gaps can be
observed in the \xx\ and \xy\ polarised spectra and that the residual
scattering below the $2\Delta_{\sigma}$ peak is due, at least in part,
to an intrinsic $\pi$ band contribution with a dirty-limit gap
\cite{Quilty03:pc385}. Nonetheless, the residual scattering seen in
\abplane\ measurements varies in both strength and character between
both polarisations and crystals \cite{Quilty02:prl88}, hindering a
precise determination of the $\pi$ band contribution to \xx\ and \xy\
spectra.

The measurements confirm two gap model calculations
\cite{Shulga01:0103154,Liu01:prl87}, which indicate a gap on the $\pi$
sheets with a magnitude approximately one third that of the $\sigma$
sheets. Furthermore, the gap magnitudes determined here fall within
the range of values measured by other experimental techniques
\cite{Buzea01:sst14}. Neither $\Delta_{\sigma}$ nor $\Delta_{\pi}$
show evidence of strong anisotropy; the measured anisotropy is of the
order of 5~\Rcm\ at most \cite{Quilty03:pc385}, consistent with the idea
that impurity scattering will strongly suppress gap anisotropy in
MgB$_{2}$ \cite{Mazin02:0212417}. Despite the satisfactory agreement
between our fits and the data (Fig.~\ref{fig:SuperFit}), some
discrepancies remain. It is possible that a more complex description
of the continuum, which accounts for effects such as a long coherence
length in the $\pi$ bands \cite{Eskildsen03:pc385}, interband scattering
effects, and Raman vertex corrections, would resolve these
discrepancies.

In conclusion, Raman scattering from the \acface\ of thick MgB$_{2}$
single crystals allows Raman scattering from the $\pi$ and $\sigma$
Fermi surfaces to be discerned. In the superconducting state a sharp
pair-breaking peak at just over 100~\Rcm\ is seen only in \xx\
polarisation and we conclude that this feature is associated with
clean-limit superconductivity on the $\sigma$ bands. In contrast, a
threshold unaccompanied by a sharp peak appears at around 29~\Rcm\ in
the \zz\ and \xz\ continuum which we determine to be a strongly damped
superconductivity-induced renormalization on the $\pi$ bands. The
Raman spectra thus successfully distinguish two gaps in MgB$_{2}$, a
larger gap on the $\sigma$-bands and a smaller dirty-limit gap on the
$\pi$-bands, with relatively strong impurity scattering restricted to
the $\pi$ band system. These observations are unusual even if we take
into account the multiple-band structure of MgB$_{2}$. A dirty-limit
gap on the $\pi$-band might arise if the dominant cause of carrier
scattering is defects in the magnesium layers, affecting only the
$\pi$-band electrons \cite{Mazin02:prl89}.

This work was supported by the New Energy and Industrial Technology
Development Organization (NEDO) as collaborative research and
development of fundamental technologies for superconductivity
applications.

\bibliography{Journal_Abbreviations,papers,MgB2_Papers}

\end{document}